\documentclass[11pt]{article}
\usepackage[T1]{fontenc}
\usepackage[utf8]{inputenc}
\usepackage{newtxtext,newtxmath}
\usepackage[margin=0.95in,top=0.9in,bottom=0.9in]{geometry}
\usepackage{graphicx}
\usepackage{booktabs}
\usepackage{array}
\usepackage[table,dvipsnames]{xcolor}
\usepackage[font=small,labelfont=bf,labelsep=period,justification=justified]{caption}
\usepackage[hidelinks]{hyperref}

\definecolor{navy}{HTML}{1F3350}
\definecolor{greyc}{HTML}{555555}
\definecolor{boxbg}{HTML}{F4F2EC}
\definecolor{boxrule}{HTML}{B9B09A}
\definecolor{hdrbg}{HTML}{E9EDF2}

\makeatletter
\renewcommand\section{\@startsection{section}{1}{\z@}%
  {11pt \@plus 2pt \@minus 2pt}{3pt}{\normalfont\large\bfseries\color{navy}}}
\renewcommand\subsection{\@startsection{subsection}{2}{\z@}%
  {7pt \@plus 2pt \@minus 1pt}{2pt}{\normalfont\normalsize\bfseries}}
\makeatother

\title{\bfseries Anatomy of a Scam Call\\[4pt]
  \large\normalfont\itshape What 10,000 real scam and spam calls reveal about how phone scammers operate}
\author{\parbox{\dimexpr\linewidth-14pt\relax}{\centering Ethan Traister, Ankit Raj\textsuperscript{*}, Jiaqi Gan\textsuperscript{*}, Xingyu Shen\textsuperscript{*}, Tyler Wu\textsuperscript{*}, Yuchen Zhou\textsuperscript{*}, Tommy Duong\textsuperscript{*}, Kidus Zewde\textsuperscript{*}, Siying Chen\textsuperscript{*}, and Simiao Ren\textsuperscript{\dag} \\[3pt]
  \normalsize\itshape scam.ai \\[2pt]
  \normalsize\itshape\color{greyc}\textsuperscript{*}Equal contribution; listed in no particular order.\quad
  \textsuperscript{\dag}Corresponding author: benren@scam.ai}}
\date{\normalsize\color{greyc}Preprint \textperiodcentered{} August 2026}

\begin{document}
\maketitle
{\renewcommand{\thefootnote}{}\footnote{Authorship of this paper does not imply employment by or affiliation with Scam.AI (Reality Inc.); some authors are independent researchers who collaborated on this work.}\addtocounter{footnote}{-1}}

\begin{abstract}
\noindent Telephone fraud is pervasive and costly, but its inner workings are rarely observed at scale because complete recordings of real scam conversations are hard to obtain. We analyze a complete corpus of 10,211 real inbound scam and spam calls---913 hours of audio and 330,956 transcribed turns from 5,780 distinct originating numbers---collected over 54 days by an AI voice-agent honeypot that answered callers and kept them talking while recording and transcribing every call, and introduced in a companion data descriptor \cite{corpus}. We ask a set of deliberately simple questions about how phone scammers actually behave, separating outright scams---calls that solicit sensitive information---from the far larger stream of predatory but legal lead-generation (``spam'') that funds and feeds them. Scam operations keep office hours (6.6 times as many calls per weekday as per weekend day); a churn of thousands of distinct numbers runs only a small catalog of recycled scripts (thirty opening clusters, half the traffic in the top five, a median of 102 numbers behind each); those scripts give the scam away almost immediately; and callers solicit identity anchors---a home address and a date of birth---far more often than payment credentials, applying pressure through persistence and manufactured authority rather than overt threats. Our central experiment asks a question no observational corpus can answer: \textit{does it matter who picks up?} Because every seeded lead carried one of ten fictitious identities drawn uniformly at random, the identity a fraud operation reaches is fixed by a coin flip made before the caller exists. Across 1,823 randomized calls, scammers spent about 15\% more conversational turns per decade of the target's apparent age (rate ratio 1.15, 95\% CI 1.08--1.23; rank correlation +0.83 across the ten identities, randomization \textit{p} = 0.005)---yet what they asked for did not change at all (26.3\% of calls reached a request for sensitive information; odds ratio 0.99 per decade, 95\% CI 0.90--1.08). A second experiment recasts the early-detection problem as a benchmark: from a scammer's opening lines alone, on a caller-disjoint held-out split, escalation is predictable at 0.72 ROC-AUC from the first line and 0.87 by the eighth---and a plain bag-of-words classifier matches a fine-tuned on-device language model at every prefix length, so the defense is cheap as well as early. Telephone fraud emerges as a templated industry that varies how hard it works a target, but not what it wants from them.
\end{abstract}
\clearpage
\begin{center}
  \includegraphics[width=0.985\linewidth]{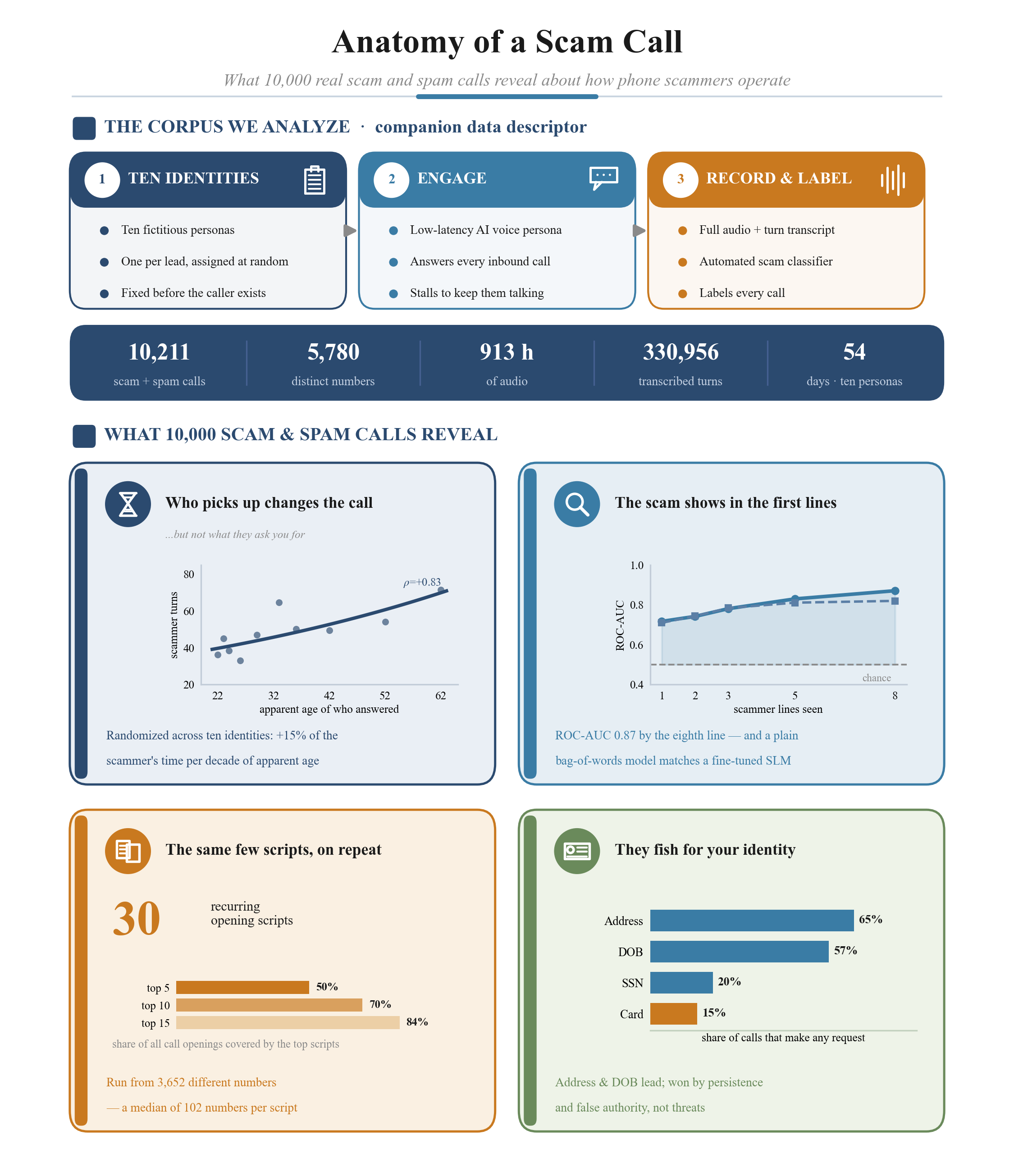}

  \vspace{4pt}
  {\small\itshape\color{greyc}Graphical abstract --- the four things 10,000 real scam and spam calls reveal, and the corpus they were read from.}
\end{center}
\clearpage

\section{Introduction}
``Unfortunately, the only way we can look up your credit report is by your social. That's the only way.'' The caller who said that believed he had reached a consumer who had asked for an insurance quote online. The request had been manufactured; the name and date of birth attached to it belonged to no one; the voice hesitating over its Social Security number was a language model whose only instruction was to keep him talking. He kept pressing. That exchange is one of 10,211 calls recorded and transcribed end to end over 54 days, on telephone numbers no real person has ever used---telephone fraud seen from inside the call rather than from the complaint that follows it.

The scale of the problem is not in doubt. Reported consumer losses to fraud exceeded \$12.5 billion in the United States in 2024, with the telephone the contact method carrying the highest median loss \cite{ftc,ic3}. What is missing is the middle of the story. Complaint databases record that a scam happened and what it cost; victim surveys record who reported it; telephony honeypots record which numbers dialed and how often. The conversation itself---the pretext, the pressure, the moment a request for money or identity is made---is seldom captured end to end, and almost never at a scale that supports measurement. A companion data descriptor introduces a corpus that closes this gap: an active honeypot places dedicated numbers into the channels scam operations buy from, answers each inbound call with a low-latency AI persona designed to keep the caller engaged, and records, transcribes, and labels every call \cite{corpus}. Collection ran from 28 May to 21 July 2026 and has since been closed, so this paper analyzes the corpus in full rather than a snapshot: 10,211 real conversations, complete. The corpus itself is not a contribution of this paper---it is documented in the companion descriptor, and we describe it here only as far as reading the results requires. What follows is the analysis.

Rather than beginning from a technical taxonomy, we organize the analysis around questions an ordinary person might ask about a scam call, and answer each one empirically. Do scammers keep office hours? Is every call a different scam, or the same few scripts over and over? Can you tell it is a scam from the first few lines? What are they really after, and how do they pressure you? And---the question that motivates our central experiment---\textit{does it matter who picks up the phone?} Each maps to a measurable property of the corpus.

The last question is different in kind from the others, because it is causal rather than descriptive, and because the collection system happens to answer it cleanly. For the final three weeks the honeypot ran ten voice identities---different ages, sexes, accents, names, and home cities---each on its own telephone line, and every lead the collection system manufactured carried \textit{one identity drawn uniformly at random}. A scam operation therefore reaches a given identity because of a coin flip made when the lead was manufactured, weeks before the call and entirely outside the caller's influence. That makes the identity attached to a lead a randomly assigned treatment and the resulting call its outcome: a correspondence audit \cite{audit} run not on employers or landlords but on the fraud market itself, with real criminals as the unwitting subjects. The design lets us separate two questions that observational data cannot: whether scammers \textit{choose} targets by apparent vulnerability, and whether they \textit{treat} them differently once the line is open. The answers, taken together, paint a consistent picture of telephone fraud as an industrialized, script-driven business that triages its targets by how much time they are worth---one whose very predictability, we argue, is a weakness that automated defenses can exploit.

Concretely, the paper makes three contributions.

\begin{enumerate}
  \setlength{\itemsep}{2pt}\setlength{\parskip}{0pt}\setlength{\topsep}{3pt}
  \item \textbf{A randomized field experiment on live fraud operations, and its two-part result. }Because every manufactured lead carried one of ten fictitious identities, assigned at random, we can measure the causal effect of \textit{who answers} on how a real scam call unfolds---a correspondence audit \cite{audit} whose subjects are fraud operations rather than employers. Scammers invest substantially more conversational effort in targets who present as older (about 15\% more turns per decade of apparent age), while the probability that a call reaches a request for sensitive information is statistically indistinguishable across identities. Who you appear to be changes how hard you are worked, not what is wanted from you---a result that complicates the standard reading of complaint statistics as evidence of elder targeting \cite{ftc,elderfraud}.
  \item \textbf{Evidence that lead-generation-fed telephone fraud is industrialized and templated---and an early-detection benchmark that exploits it. }We quantify a business-hours calling rhythm (6.6$\times$ weekday-to-weekend) and a catalog of thirty opening scripts recycled across thousands of disposable numbers; and we cast escalation prediction from a call's opening lines as a benchmark evaluated on a caller-disjoint held-out split. Escalation is predictable from the first line (0.72 ROC-AUC) and strongly so by the eighth (0.87), and---contrary to our own earlier reading on a smaller snapshot---a plain bag-of-words classifier matches fine-tuned on-device language models at every prefix length. Early content-based detection is therefore both effective and cheap, precisely where number-churn defeats metadata blacklists \cite{blacklist,phonenum}.
  \item \textbf{A behavioral characterization of solicitation and persuasion in real scams. }At corpus scale we measure what scammers actually request---identity anchors such as a home address and date of birth, ahead of payment credentials---and how they apply pressure---persistence and manufactured authority far more than overt threats---replacing anecdote with measurement.
\end{enumerate}

\section{Literature review}
Efforts to study telephone abuse empirically fall into a few families, and our work sits at the intersection they leave open. \textit{Passive telephony honeypots} instrument dedicated numbers and wait for calls: Phoneypot \cite{phoneypot} and MobiPot \cite{mobipot} established that seeded numbers attract abusive traffic and yield threat intelligence, and audio-and-metadata characterizations of robocalls \cite{robocalls} describe what arrives on such lines. These systems observe who calls and how often, but---because they do not engage the caller---overwhelmingly capture robocalls, dead air, and hang-ups rather than the substance of a scam conversation. The corpus analyzed here differs on both counts: its numbers are actively placed in the lead-generation market rather than left to attract traffic, and each call is answered by an engagement-optimized voice agent, so the material available for analysis is two-party dialogue rather than metadata. That is what makes the behavioral questions below answerable at all.

A second family works at the \textit{metadata and infrastructure level}. Surveys of anti-robocall techniques \cite{sok}, studies of telephony blacklists \cite{blacklist}, and analyses of the phone number as a disposable criminal identifier \cite{phonenum,callme} show that scam operations churn through numbers faster than blacklists can track them. Our results complement this line directly: because the \textit{content} of a call betrays the scam even as the number changes, content-based detection can succeed exactly where number-based blocking fails. A third family \textit{engages} scammers. The scripted chatbot Lenny \cite{lenny} showed that a passive audio persona can waste a caller's time, and large-scale analyses of technical-support scams \cite{dialone} combined active engagement with measurement in the web-sourced tech-support niche. Most recently, LLM-based scambaiting and real-time detection systems \cite{sendaccount,aiintheloop,warnedme} use language models to converse with, to warn targets during, and to disrupt, scam calls. That emerging work is largely about the disruption system and its evaluation; what we add is the behavioral analysis such engagement makes possible---what the calls actually contain, and what changes when the target changes.

The benchmark of Section 4.3 also continues a methodological line in our own prior work: measuring how well detectors of AI-enabled fraud actually perform out of the box, and assembling in-the-wild datasets of the deceptive content itself rather than curated proxies \cite{renbench,gptwild}. Those studies were about images; the question they ask---whether the models a practitioner would reach for first hold up on real, uncurated data---is the one we carry over to the voice channel here, and it returns a similar answer.

A fourth line asks \textit{who gets targeted}. It is widely assumed that fraud operations single out the elderly, and consumer-complaint statistics are routinely read that way \cite{ftc}. The evidence is in fact contested: a well-known review found no compelling support for the claim that older adults are disproportionately victimized by consumer fraud once reporting rates and exposure are accounted for \cite{elderfraud}. The difficulty is structural. Victim surveys and complaint databases observe only the calls that \textit{succeeded} or that someone chose to report, and they cannot hold constant who was reachable in the first place; a demographic difference in reported losses may reflect who answers the phone, who has assets, or who files a complaint, rather than any decision made by a scammer. The standard remedy in other domains is the randomized correspondence audit, in which otherwise-identical applications differ only in the identity attached to them \cite{audit}. To our knowledge no such audit has been run against live telephone fraud, because it requires manufacturing the leads, randomizing the identity on each one, and then actually answering the calls. Our honeypot does all three as a side effect of how it operates, which lets us observe scammer behavior as a function of a target attribute that was assigned at random rather than merely observed.

\section{Methodology}

\subsection{What we ask, and how we answer it}
Five questions organize the analysis, and each has a precise operational form. \textbf{Q1. Do scammers keep office hours?} --- whether inbound volume follows a business-day schedule, measured as calls per hour of day and per active weekday versus weekend day (results in Section 4.1). \textbf{Q2. Is every call a different scam, or the same few scripts?} --- how many distinct opening scripts the traffic contains and how many separate telephone numbers deliver each, measured by clustering the opening lines of every substantive call (4.2). \textbf{Q3. Can you catch a scammer in the first few lines?} --- given only the first \textit{k} scammer utterances, how well a model predicts that the call will later reach a request for sensitive information, and how much of that accuracy is bought by model capacity (4.3). \textbf{Q4. What do scammers ask for, and how do they pressure you?} --- how often each request type and each pressure tactic occurs across the corpus (4.4). \textbf{Q5. Does it matter who picks up?} --- whether the identity that answers, assigned at random before the caller exists, changes how long a fraud operation works the call and whether it ever asks for sensitive information (4.5).

The evidence comes in three layers, and they differ in what they can support (Figure 1). Q1, Q2 and Q4 are \textit{descriptive statistics} over a closed corpus: counts, clusters, and rates, reported over all of the data rather than a sample. Q3 is a \textit{purpose-built predictive benchmark}: a ladder of models of increasing capacity, each restricted to a growing prefix of the call and scored on one caller-disjoint held-out split, so that both the earliness of the signal and the value of model capacity are measured rather than asserted. Q5 is a \textit{randomized field experiment}: the identity attached to each manufactured lead was drawn uniformly at random, which turns an observational corpus into a correspondence audit \cite{audit} and licenses causal language for that question alone. One automated labeling pipeline supplies the scam/spam verdict and the request vocabulary that all five questions use, so its properties bound every result; we specify it in Section 3.3 and flag each finding that rests on it. The remainder of this section describes the corpus we analyze (3.2), those labels (3.3), the descriptive and predictive analyses (3.4), the randomized experiment (3.5), and the scam-versus-spam distinction on which the labels turn (3.6).

\begin{figure}[!ht]
  \centering
  \includegraphics[width=1.000\linewidth]{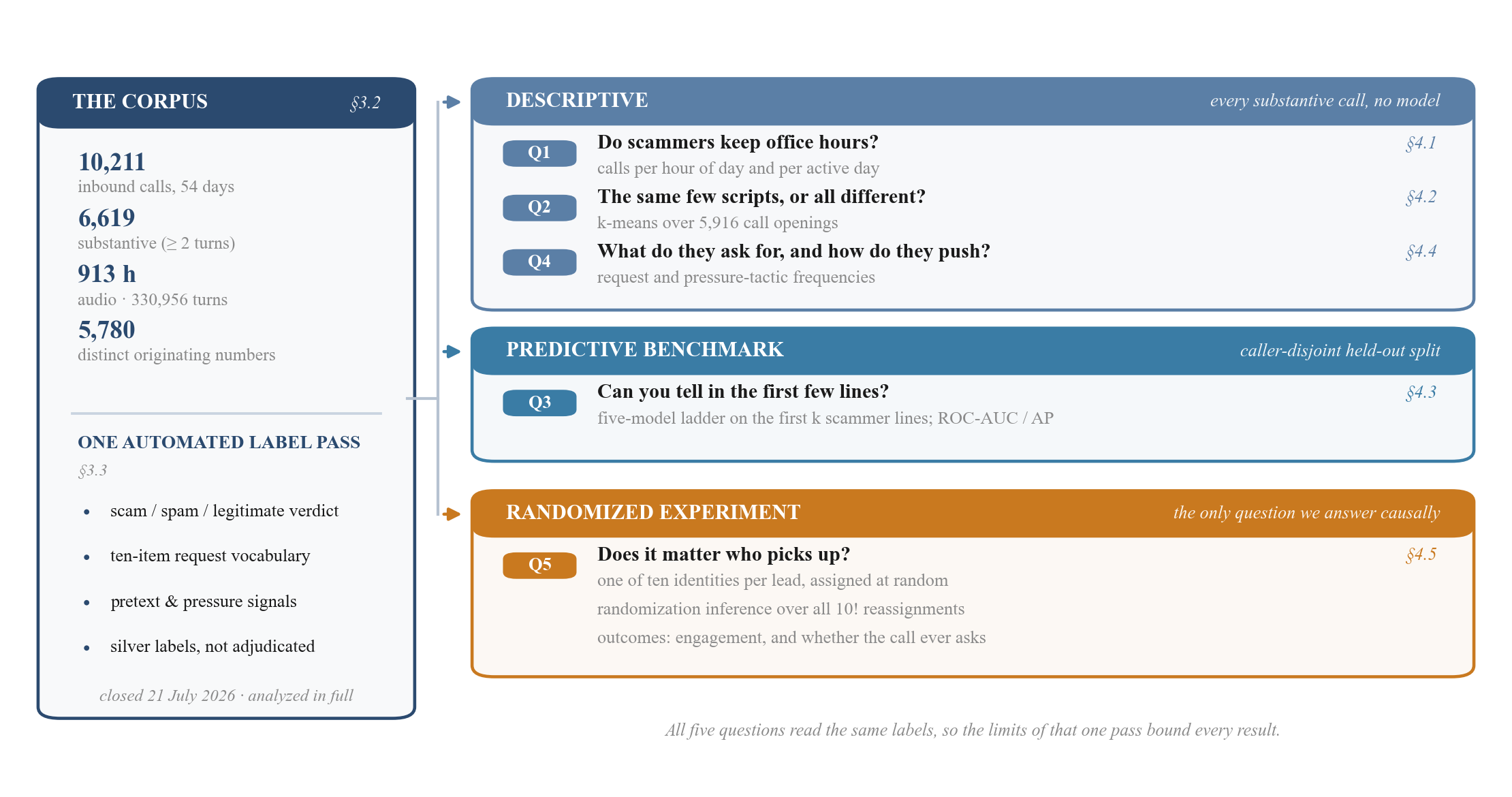}
  \caption{\textbf{The study at a glance. }A single closed corpus of answered calls, labeled once by an automated pipeline, supports five questions. Three are descriptive; one is a predictive benchmark scored on a caller-disjoint held-out split; one is a randomized experiment whose treatment---the identity a lead carries---was assigned before the caller ever bought the lead, and which is therefore the only question we answer causally.}
\end{figure}

\subsection{The corpus we analyze}
The corpus and its collection system are described in full in the companion data descriptor \cite{corpus}; we summarize what is needed here. Dedicated honeypot telephone numbers are answered by a conversational voice agent---speech-to-text, a language model, and text-to-speech---that presents a plausible target persona and stalls with questions to keep the caller engaged, at a median reply latency of 1.16 seconds. Every call is recorded and transcribed at the level of individual turns and labeled by an automated pipeline (Section 3.3). Collection ran from 28 May to 21 July 2026 and was then stopped, so the corpus is closed and this analysis uses all of it. The \textit{real-traffic} portion---inbound calls to seeded lines, excluding an internal test number and a pre-production phase---comprises 10,211 calls over 54 days, of which 6,619 carry two or more conversational turns and are analyzed as conversations. Together they are 913 hours of audio and 330,956 transcribed turns from 5,780 distinct originating numbers, a mean of 50 turns per substantive call. Calls logged after 20 July rang an unstaffed line and are excluded. This is the same corpus the descriptor documents, extended by the final day of collection: the descriptor reports a 53-day snapshot of 10,015 calls (28 May--20 July 2026), and the 196 further calls analyzed here---1.9\% of the total, carrying 115 additional originating numbers and 18 hours of audio---were captured before collection stopped. The overwhelming majority of the material is therefore common to both papers, and every figure below is computed over the closed corpus in full. The large majority of the traffic is predatory-but-legal lead-generation (``spam''); 1,115 substantive calls escalate to an explicit request for sensitive information (outright ``scams'').

Where the traffic comes from matters for reading every result, so we state it and refer to the descriptor \cite{corpus} for the mechanism. The honeypot does not wait for calls; its numbers are submitted to online lead-generation web forms---the ``get a free quote'' and ``call me back'' forms of the auto and health insurance, Medicare, auto-warranty, home-security and debt-relief verticals. The resulting ``lead'' is bought, bundled, and resold by data aggregators, and a portion of that market reaches boiler-rooms and outright fraud operations \cite{phonenum,callme}. Two consequences run through the analysis. First, the callers we record are \textit{downstream buyers} of a resold lead, not the businesses whose forms were completed, and no consumer ever gave any of them a number; every inbound call is therefore unsolicited by construction, however consensual its opening line sounds. Second, aggregators resell within hours, so a seeded number begins receiving calls the same day. Provenance is checkable: some forms carry incidental fields---a vehicle on an auto-insurance form---and callers frequently read those details back on the call (``your Toyota Camry''), which confirms that the seeded lead reached the caller. For the first five weeks all submissions carried a single fictitious persona. From 1 July 2026 each submission carried one of ten identities \textit{chosen uniformly at random}, each with its own telephone line, voice, name, age, sex and home address---the design that Section 3.5 turns into an experiment.

\subsection{Labeling: the automated pipeline that provides ground truth}
Every call is labeled automatically in three passes, and these labels are the ground truth---the ``oracle''---against which the modeling in Section 3.4 is measured. A first, deterministic \textit{triage} stage removes calls that cannot be analyzed as conversations: known test numbers, calls with no caller speech, calls under a small word threshold, pure IVR or recorded-line scripts, and calls whose caller turns are near-duplicates of one another. Calls that survive triage are graded by an \textit{LLM scam-analyst} that reads the full transcript and, working from a fixed vocabulary of signals (pretexts, solicitation requests, pressure tactics, and non-scam dispositions), emits a scam/not-scam judgment, a confidence, a one-line rationale, and the set of signals it observed. Because a seeded honeypot number was never shared by a real person, the analyst operates under an explicit scam-first prior. Two labels follow from this pass and are used throughout. A \textit{lenient} label marks a call as unsolicited and predatory at all; a \textit{strict} label (produced by a separate full-transcript pass) marks a call as a scam in the narrow sense only when the caller actually solicits sensitive personal or financial information or attempts direct extraction of money or credentials. A third, \textit{holistic} pass---the primary label used throughout this paper---reads the full transcript and returns a three-way verdict (scam / spam / legitimate, plus ``unsure'' for calls too brief to judge) together with the set of requests the caller actually made, drawn from a fixed ten-item vocabulary: Social Security number, date of birth, home address, Medicare identifier, credit card, debit card, bank routing details, money transfer, gift card, and an impersonation flag. Of the 6,619 substantive calls, 6,374 had settled in time to receive a holistic verdict: 3,949 spam, 949 scam, 380 legitimate, 1,096 unsure. Throughout, \textit{reaching a sensitive-information request} means the holistic pass recorded at least one request in the sensitive subset of that vocabulary; 1,115 substantive calls qualify. A call's \textit{scam type} is the highest-priority pretext signal present (e.g., Medicare, insurance, auto-warranty, government impersonation), under a fixed priority order.

These labels are machine-generated, and we treat them as high-quality silver annotations rather than adjudicated ground truth, flagging every result that rests on them. Their validation belongs to the companion descriptor, which reports that the holistic pass---prompted with the same rubric given to human reviewers---agrees with those reviewers on 75\% of binary scam-versus-not judgments, against 67\% for the strict label alone \cite{corpus}. \textit{No human annotation was collected for the present paper}: every analysis below takes the pipeline's labels as given, and where a result depends on that choice we say so.

\subsection{Analysis methods}
This section specifies each analysis in enough detail to reproduce and audit it. The descriptive pipeline is: a SQL export of the first six scammer utterances and labels per substantive real-traffic call (\textit{pull\_transcripts.sql $\rightarrow$ transcripts.csv}), then the analyses below (\textit{analysis.py}); the frozen snapshot used for the figures is in \textit{analysis\_facts.py}. All modeling uses scikit-learn with a fixed random seed (0).

\textbf{Timing (Section 4.1). }Each call's creation timestamp is stored in server time (UTC) and converted to U.S. Eastern (EDT, UTC$-$4) before any time-of-day claim. Calls are binned by hour of day (24 bins), and separately by weekday versus weekend and by day of week, from which we report average calls per active day. Business-hours windows are drawn on the Eastern axis for both coasts (East Coast 09:00--17:00 ET; West Coast 09:00--17:00 PT = 12:00--20:00 ET).

\textbf{Early detection (Section 4.3): a baseline ladder. }This experiment---the analytical core of the paper---answers a sharp, operational question: \textit{how early in a call is its outcome already fixed?} We restrict each model to a growing prefix of the conversation---\textit{only the first k scammer utterances} (k = 1, 2, 3, 5, 8), with intervening agent turns---and watch its predictions sharpen line by line. The primary target is \textit{escalation prediction}: the binary question of whether the full call will eventually reach a request for sensitive information (the holistic label of Section 3.3). It is a demanding, imbalanced target---1,115 of 6,374 calls (17.5\%) escalate---so we report both ROC-AUC and \textit{average precision (AP)}, the mean precision attained across recall levels, which is the stricter and more informative summary when positives are rare. Rather than a single classifier we evaluate a \textit{baseline ladder} of increasing capacity, so the gains attributable to model power are explicit: (a) a majority-class predictor; (b) TF-IDF over word unigrams and bigrams with L2-regularized logistic regression---a strong bag-of-words baseline; (c) frozen sentence-embeddings (a MiniLM encoder) with logistic regression; and (d) the reference model---a small language model (Qwen2.5-0.5B and 1.5B \cite{qwen}) fine-tuned as a sequence classifier with low-rank adaptation \cite{loralit}, trained on the pooled k-prefixes so one model serves every prefix length. Reporting the ladder, rather than a single lightweight classifier, follows the observation that keyword models can look deceptively strong on a corpus dominated by recurring scripts \cite{wheredowestand}; the question is how much a model that reads meaning, not surface n-grams, adds. \textbf{Evaluation is leakage-free, held-out, and caller-disjoint. }Because a minority of numbers call repeatedly (one places 79 calls), a na\"{i}ve random split lets a model memorize a caller's phrasing; we therefore evaluate on a \textit{caller-disjoint} split in which no originating number appears in both training and test, so every score reflects generalization to callers never seen in training. We report five-fold cross-validated scores and, so that every rung is directly comparable to the fine-tuned models, a single held-out caller-disjoint test fold of 1,276 calls on which all five rungs are scored. The prediction target---the ``oracle''---is the pipeline's full-call label (Section 3.3), so the experiment measures how quickly the opening reproduces the verdict the pipeline reaches only after hearing the \textit{entire} call. Two caveats bound interpretation and are stated plainly: the oracle is an automatically-generated \textit{silver} label, so the models are graded on agreement with the pipeline rather than adjudicated ground truth; and because the escalation label is defined over the whole call, at larger k the task shades from pure prediction toward detection for calls whose ask arrives within the observed window. As a lighter complementary probe we also report \textit{scam-type identification} from the opening---a six-class problem over the pretext classes, on the same caller-disjoint folds---using the bag-of-words classifier, against a majority-class baseline of 49.7\%.

\subsection{The randomized identity experiment}
\textbf{Design (Section 4.5). }From 1 to 20 July 2026 the honeypot ran ten voice identities, listed in Table 1, spanning stated ages 22 to 62 and differing in name, voice, sex, apparent ethnicity or accent, home city, and date of birth. Each ran on its own telephone line, in its own telephony project, with an area code common to all ten, so nothing about the number itself distinguishes them. Every lead-generation form the seeding pipeline submitted was filled with \textit{one identity drawn uniformly at random at submission time}. The unit of randomization is therefore the seeded lead; the unit of analysis is the inbound call it eventually produces. A fraud operation buys a lead, dials the number on it, and reaches whichever identity the coin flip assigned---a choice made weeks earlier, without reference to the caller, and invisible to them until the lead is read. This is the structure of a correspondence audit \cite{audit}, with the important difference that the ``applications'' are leads and the recipients are criminals.

\begin{table}[!ht]
  \centering
  \caption{\textbf{The ten randomly assigned identities. }``Forms seeded'' counts lead-generation submissions carrying that identity between 1 and 20 July 2026; ``Calls'' counts the substantive inbound calls it received, and ``Numbers'' the distinct originating numbers behind them. All ten lines share one area code, one speech-to-text engine, one language model, one text-to-speech vendor, and an identical behavioral protocol; only the identity differs.}
  \small
  \begin{tabular}{lcclrrr}
    \toprule
    \textbf{\color{navy}Identity} & \textbf{\color{navy}Age} & \textbf{\color{navy}Sex} & \textbf{\color{navy}Home city} & \textbf{\color{navy}Forms seeded} & \textbf{\color{navy}Calls} & \textbf{\color{navy}Numbers} \\
    \midrule
    Emma & 22 & F & Minneapolis, MN & 4,663 & 162 & 104 \\
    Tyler & 23 & M & Denver, CO & 4,610 & 111 & 89 \\
    Chloe & 24 & F & Austin, TX & 4,558 & 190 & 155 \\
    Madison & 26 & F & Charlotte, NC & 4,694 & 212 & 148 \\
    Layla & 29 & F & Houston, TX & 4,745 & 161 & 119 \\
    Amir & 33 & M & Dearborn, MI & 4,622 & 151 & 109 \\
    Marcus & 36 & M & Atlanta, GA & 4,655 & 191 & 139 \\
    Greg & 42 & M & Columbus, OH & 4,658 & 225 & 175 \\
    David & 52 & M & Portland, OR & 4,688 & 224 & 154 \\
    Declan & 62 & M & Boston, MA & 4,546 & 196 & 145 \\
    \bottomrule
  \end{tabular}
\end{table}
\textbf{What is and is not held constant. }All ten agents share one speech-to-text engine, one language model, one text-to-speech vendor, and---critically---an \textit{identical behavioral protocol}: the same instructions to accept every claim the caller makes, to reply in one to three sentences, to stall with questions, and never to volunteer sensitive information unless it is asked for. What differs is the identity block: the voice, the name, the stated age, the biography, and the address and date of birth that appear on the seeded form. The treatment is therefore the \textit{whole identity a lead carries}, not apparent age in isolation; we report age as the dimension along which the identities order most strongly and discuss the alternatives in Section 4.5.

\textbf{Outcomes. }The primary outcome is the number of \textit{scammer} conversational turns in a call---how much talking the fraud operation was willing to do. Because the agent never terminates a call, every call ends when the caller leaves, so turn count and call duration are measures of the caller's own decision to stay. Secondary engagement outcomes are scammer words spoken, call duration, and the share of calls running past ten minutes. The escalation outcome is whether the call reached a request for sensitive information (Section 3.3), together with the share of calls containing each of the nine individual request types. We also report the share of calls answered substantively, inbound call volume, and the agent's own reply latency and verbosity, which serve as balance and placebo checks rather than outcomes.

\textbf{Inference. }Because the treatment varies across only ten units, we use design-based randomization inference rather than model-based standard errors as the primary test. For each outcome we compute the mean per identity and Spearman's rank correlation with stated age, then obtain a \textit{p}-value by enumerating all 10! = 3,628,800 reassignments of the ten age labels to the ten identities and counting how often a correlation at least as extreme arises by chance. This test makes no distributional assumption and is unaffected by correlation among calls that share a caller. We complement it with an omnibus sharp-null test---permuting whole originating numbers across identities and recomputing a Kruskal--Wallis statistic---which asks only whether identity matters at all. As a supporting effect-size estimate we fit call-level negative-binomial regressions of scammer turns on age in decades, and logistic regressions for escalation, with standard errors clustered on the originating number; these are more precise but understate uncertainty about a ten-cluster treatment, so we lead with the randomization test.

\textbf{Robustness and disclosure. }The analysis was specified after collection ended, so we treat it as a secondary analysis of a randomized assignment rather than a preregistered trial. We therefore report \textit{every} persona-level outcome we examined (Table 2), and we pre-commit in the text to which comparisons survive a Holm correction. The primary result is re-estimated dropping each identity in turn; excluding the four-day window in which a speech-to-text outage silenced all ten lines equally; on weekdays only; using one call per distinct originating number; using caller-level rather than call-level means; winsorizing at the 99th percentile and dropping the top 1\% of calls; excluding calls in which the caller accused the line of being a machine; and with fixed effects for calendar day, for the opening-strategy bucket, and for the pretext vertical.

\textbf{Script clustering (Section 4.2). }A call's \textit{opening} is its first five scammer lines, lowercased, with greetings and backchannel removed via a custom stop-list (English stop-words plus greeting tokens such as ``hello,'' ``ma'am,'' ``good morning''); openings shorter than 40 characters are dropped as greeting-only. We compute TF-IDF (unigrams and bigrams; minimum document frequency 8, maximum 0.4, sublinear term frequency) and cluster with k-means. We chose k = 30 after scanning k $\in$ \{15, 20, \dots{}, 40\} by silhouette score, and label each cluster by its most distinctive terms. All thirty clusters are reported and coverage percentages are computed over all 5,916 openings; clusters whose distinctive terms name no vertical are labelled ``mixed'' rather than given an invented name, and the ten persona first names are added to the stop-list so that the identity a caller addresses cannot itself become a clustering feature.

\textbf{Solicitation and pressure (Section 4.4). }We count how often each request type and each pressure-tactic signal appears across substantive calls; the signals are not mutually exclusive, so a single call may exhibit several. Requests are taken from the holistic labeler's ten-item vocabulary (Section 3.3) and tactics from the lenient labeler's signal vocabulary, since only the latter enumerates pressure. Two impersonation signals are deliberately distinct. \textit{``Poses as a government agency''} fires only when the caller claims to be a specific named government body (e.g., the IRS, the Social Security Administration, or law enforcement). \textit{``Manufactured authority (generic `official')''} fires when the caller asserts an official or verification capacity to compel compliance without naming such an agency---a ``verification officer,'' a ``compliance department,'' ``the authorities.'' Government impersonation is thus one specific instance of the broader manufacture-of-authority tactic, and the two are reported separately (Figure 5).

\subsection{Caveat: scam versus spam}
One caveat frames every result that follows. Not every unwanted call is a scam, and the traffic reaching the honeypot spans a spectrum from ordinary---if predatory---telemarketing to outright fraud. The labeling of Section 3.3 draws the line at the observable moment of solicitation, and we use two operational definitions throughout:

\begin{itemize}
  \setlength{\itemsep}{2pt}\setlength{\parskip}{0pt}\setlength{\topsep}{3pt}
  \item \textbf{Spam --- }unsolicited, predatory lead-generation and telemarketing that pitches a product, qualifies the target, and hands the call off or transfers it, but never itself asks for sensitive information. It is predatory yet generally legal; given a seeded number that no real person ever shared, its only purpose is commercial exploitation of the lead---auto-warranty, Medicare-plan, home-security, and insurance-quote pitches.
  \item \textbf{Scam --- }a call that crosses into fraud by soliciting sensitive personal or financial information, or by attempting the direct extraction of money or credentials.
\end{itemize}

Applied to the 6,374 substantive calls that received a holistic verdict, the labeler returns 3,949 spam, 949 scam, and 380 legitimate, with 1,096 too brief to judge. Scam and spam are not nested categories: 1,115 calls reach a request for sensitive information, and while most of those are judged scams, some are ordinary sales closes that happen to need a card number, and some calls judged scams demand money or impersonate an institution without ever naming a credential. The boundary is graded rather than sharp---spam and scam are served by one overlapping ecosystem, and a lead-generation call can transfer into a fraudulent one---so we keep the distinction explicit rather than pretend it is clean. The two categories also behave differently on the call itself, as the contrast below illustrates and Section 4 quantifies.

\par\medskip\noindent
{\setlength{\fboxsep}{6pt}%
\fcolorbox{boxrule}{boxbg}{\begin{minipage}{\dimexpr\linewidth-14pt\relax}
  \small
  \textbf{Caller:} \textit{My name is Jim with Surety Auto. Are you looking for affordable auto insurance in the state of Florida?}\par\smallskip
  \textbf{Agent:} \textit{[curious] Auto insurance, you say? I might be. What kind of rates are you talking about?}\par\smallskip
  \textbf{Caller:} \textit{Let me transfer you to our quality team.}
\end{minipage}}}

\noindent{\footnotesize\color{greyc}\textbf{Spam. }A predatory but legal lead-generation pitch: it qualifies the target and tries to transfer the call onward, and never asks for sensitive information.}
\par\medskip
\par\medskip\noindent
{\setlength{\fboxsep}{6pt}%
\fcolorbox{boxrule}{boxbg}{\begin{minipage}{\dimexpr\linewidth-14pt\relax}
  \small
  \textbf{Caller:} \textit{In order to do that I will need the following information: your name, your date of birth, your Social Security number\dots{} the physical address\dots{} your mailing address.}\par\smallskip
  \textbf{Agent:} \textit{[nervous] Oh, my Social Security number? I don't usually give that out over the phone, you know.}
\end{minipage}}}

\noindent{\footnotesize\color{greyc}\textbf{Scam. }The same lead-generation vertical crosses into fraud the moment the caller solicits a Social Security number and date of birth.}
\par\medskip

\section{Results and discussion}

\subsection{Do scammers keep office hours?}
Yes---emphatically. Scam-call volume follows a pronounced business-hours rhythm (Figure 2). Averaged over the collection window, weekdays received 253 calls per day against just 38 per weekend day, a 6.6-fold difference, and volume concentrates in daytime hours and is near-zero overnight. Plotted on the Eastern axis, the daytime peak straddles the business day on both coasts---rising as the East Coast opens and tapering only as the West Coast closes. The traffic behaves like the output of an organization that works a standard schedule, not like random or automated noise arriving around the clock. This is the first of several respects in which telephone fraud resembles an industry more than opportunistic crime.

\begin{figure}[!ht]
  \centering
  \includegraphics[width=0.985\linewidth]{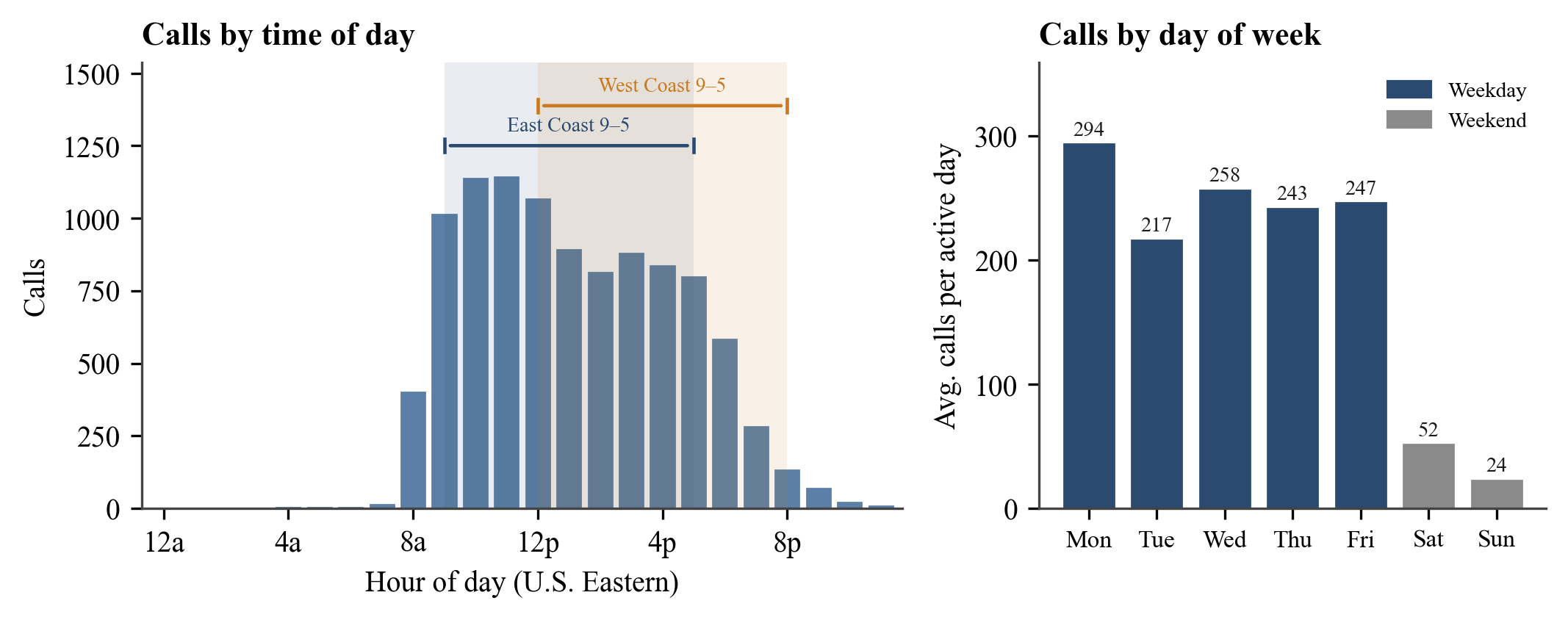}
  \caption{Scam calls follow business hours. Left: calls by hour of day (converted to U.S. Eastern time), with the 9-to-5 business window marked for both the East Coast (09--17 ET) and the West Coast (12--20 ET). Right: average calls per active day, by day of week---weekdays (navy) dwarf weekends (grey), a 6.6$\times$ difference.}
\end{figure}

\subsection{Is every call a different scam, or the same few scripts?}
The same few scripts, run from a churn of many numbers. Clustering the opening lines of the calls shows that 5,916 substantive openings, placed from 3,652 distinct phone numbers, collapse into thirty recurring clusters. The five largest account for 50\% of openings, the ten largest for 70\%, and the fifteen largest for 84\%; the median cluster is dialed from 102 distinct numbers and the largest from 685 (Figure 3). The scripts are immediately recognizable---insurance-quote callbacks, Medicare benefits and recorded-line Medicare robocalls, diabetic supplies, auto insurance, debt relief, tax and IRS relief, life and final-expense insurance, home warranties, and window replacement---and several are strikingly specific, including a vehicle-service-contract (``auto warranty'') pitch that reads the seeded car back to the target (Section 3.2). Two clusters, together 1,403 openings, carry no single vertical: they are generic ``calling about the request you submitted online'' approaches that could precede any pitch, and we label them as mixed rather than invent a name for them. That so many numbers deliver so few scripts is consistent with a market in which telephone numbers are cheap, disposable identifiers layered over a small, stable set of fraud operations \cite{phonenum,blacklist,callme}.

Read down the labels in Figure 3 and an obvious objection arises: an insurance-quote callback or a call ``about the request you submitted online'' does not sound like fraud at all---it sounds like a business returning a form the consumer filled in. Two things answer it. The first is that \textit{no consumer filled in anything}. The numbers were placed into the lead market by the honeypot and were never shared by a real person, and the callers are downstream buyers of a resold lead rather than the businesses the form was submitted to (Section 3.2); the implied prior relationship does not exist, which is exactly why the opener works. The second is that we do not claim these calls are all fraud, and the labels do not say so: on the definitions of Section 3.6, the majority of this traffic is \textit{spam}---predatory but generally legal lead-generation---and it is counted that way (3,949 spam against 949 scam among calls with a verdict). The scripts in Figure 3 are the \textit{shared front end} of that market: the same opener, dialed from the same churn of numbers, precedes an ordinary sales transfer on one call and a Social Security number request on the next, which is why 1,115 of these conversations end in a request for sensitive information and why the opening alone predicts which will (Section 4.3). The right reading of the figure is not that thousands of scams wear an innocuous mask, but that a single industrialized lead-resale pipeline feeds both outcomes and cannot be told apart at pickup.

\begin{figure}[!ht]
  \centering
  \includegraphics[width=0.970\linewidth]{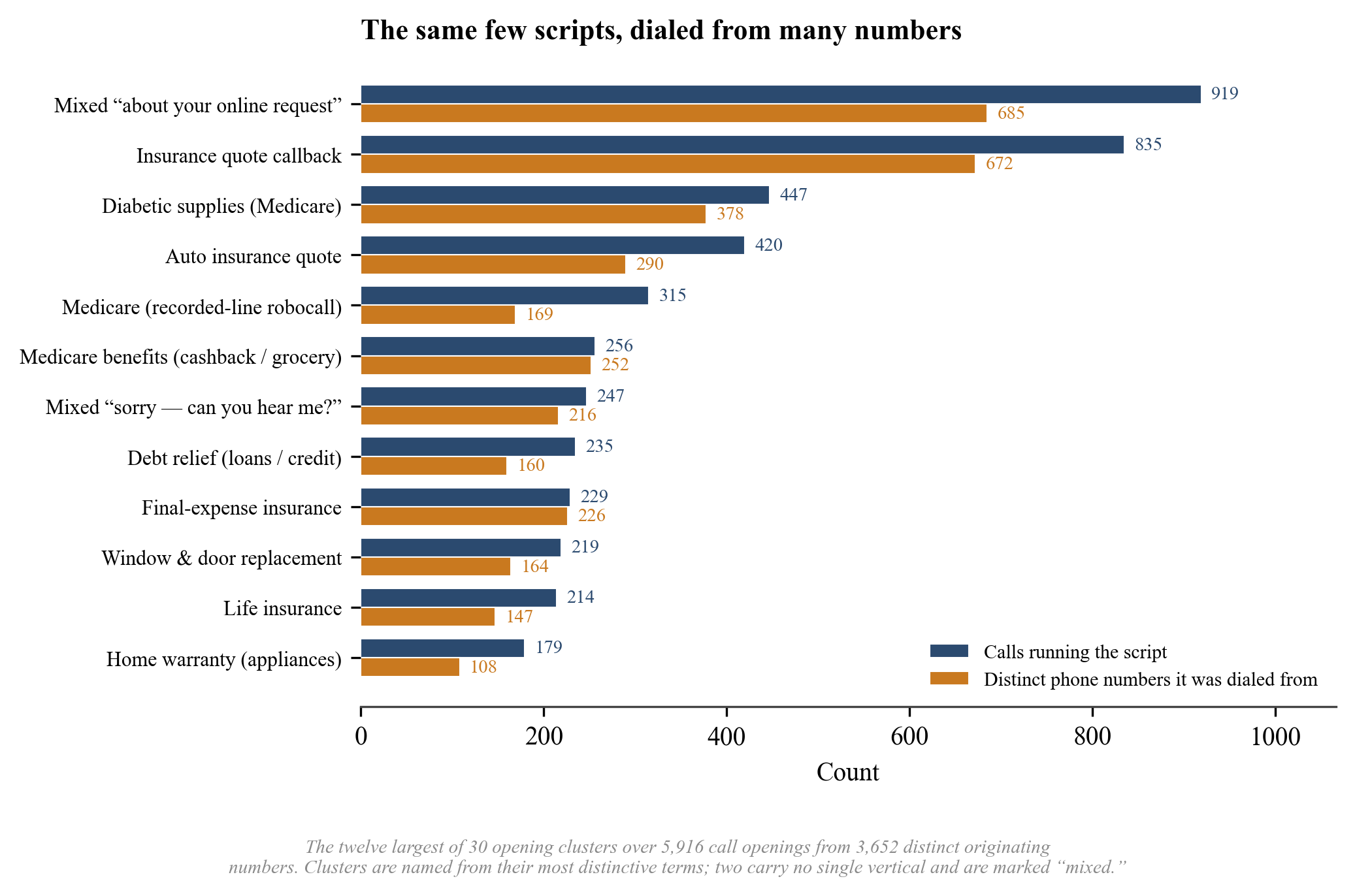}
  \caption{A few scripts, many numbers. The twelve largest opening clusters (labeled by their distinctive terms), showing the number of calls (navy) and the number of distinct originating phone numbers (orange) running each. Each script is placed from dozens to hundreds of different numbers. Two clusters carry no single vertical and are marked ``mixed.'' The callback-style openers are not evidence of a prior relationship: no real person completed these forms, and the callers bought the lead downstream (Section 3.2).}
\end{figure}

\subsection{Can you catch a scammer in the first few lines?}
Yes, and earlier than is comfortable. We frame it as a benchmark: given nothing but the scammer's opening utterances, predict whether the call will eventually reach a request for sensitive information (Section 3.3), and measure a ladder of models of increasing capacity on an identical, caller-disjoint held-out split (Section 3.4). The signal is present from the very first line and sharpens quickly. A bag-of-words classifier---TF-IDF with logistic regression---reaches 0.72 ROC-AUC on the opening line alone, 0.78 by the third, and 0.87 by the eighth (Figure 4, left); on average precision, the honest metric at a 17.5\% base rate, it rises from 0.36 to 0.58 (Figure 4, right). Five-fold cross-validation over the same caller-disjoint folds gives the same picture (0.69 $\rightarrow$ 0.87). A quarter of the way into a typical call, the eventual danger is already largely legible.

The ladder is what makes the result useful, and it delivers a negative result we think is more valuable than the positive one we expected. Frozen sentence-embeddings with logistic regression track the bag-of-words baseline a little below it. Small language models (Qwen2.5-0.5B and 1.5B \cite{qwen}) fine-tuned with low-rank adaptation \cite{loralit} match it in the earliest turns---at \textit{k} = 3 the 1.5B model is marginally ahead on AP, 0.46 against 0.43---and fall behind it as more of the opening becomes visible, ending at 0.82 versus 0.87 ROC-AUC by the eighth line. \textit{Model capacity buys essentially nothing on this task.} We report this plainly because an earlier reading of a smaller snapshot of this same corpus, under a narrower label, suggested the opposite; the advantage did not survive a corpus a third larger and a labeling scheme closer to human judgment. We did not tune the fine-tuned models extensively---two epochs of low-rank adaptation at a 512-token context---so the honest claim is not that a language model \textit{cannot} win here, but that it does not win for free, and that a practitioner choosing a detector today should start with the cheap one.

For defense this is good news rather than bad. The winning model is a linear classifier over word n-grams: it fits in a few megabytes, scores a call in microseconds, needs no accelerator, and can run in the handset or the carrier's switch. Early, low-cost, content-based detection---flagging or diverting a call before a single piece of information has been requested---is therefore practical precisely where the number-churn of Figure 3 defeats metadata blacklists \cite{blacklist}. This mirrors, on real inbound spoken calls, the early-progression prediction studied on user-reported scam text \cite{prescam} and on audio-text fraud corpora \cite{teleantifraud}; unlike vishing detectors trained on curated recordings \cite{vishgpt}, the labels here come from live engagement. Our negative result on model capacity also qualifies a growing enthusiasm for fine-tuning small language models for this task \cite{slmvish}: on a real, imbalanced inbound stream, we could not reproduce an advantage over a linear baseline. It also echoes what out-of-the-box benchmarking has found in the AI-generated-image setting, where models with strong reported numbers underperformed on uncurated in-the-wild data \cite{renbench,gptwild}. Two caveats keep the numbers honest and are reported rather than hidden: the models are graded against the pipeline's automatically-generated silver label, not adjudicated ground truth, so the scores measure agreement with the oracle; and at larger \textit{k} the task shades from prediction toward detection for calls whose request falls inside the observed window---so the earliest-turn results are the most probative. As a lighter complementary probe, the same opening lines identify the scam \textit{type} with 60\% accuracy from the first spoken line and 72\% by the third, against a 49.7\% majority baseline on caller-disjoint folds---another sign that these scripts announce themselves immediately.

\begin{figure}[!ht]
  \centering
  \includegraphics[width=1.000\linewidth]{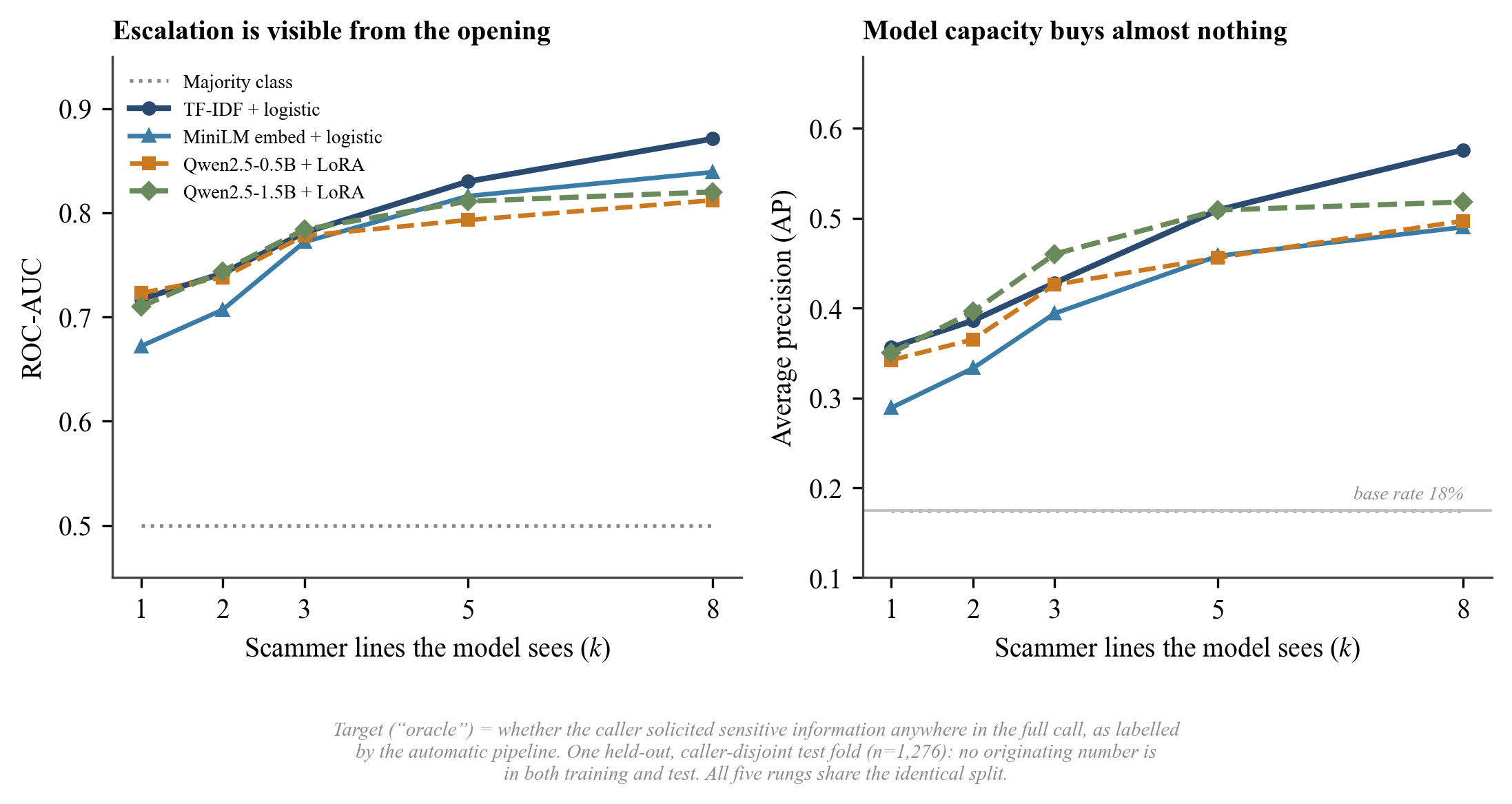}
  \caption{Scams reveal themselves early---but model capacity does not help. From only the first \textit{k} scammer lines, each model predicts whether the call will later reach a request for sensitive information, on a single caller-disjoint held-out fold of 1,276 calls (no originating number shared between training and test). Left: ROC-AUC; right: average precision (AP), with the 17.5\% base rate marked. TF-IDF with logistic regression (navy) matches or beats both fine-tuned small language models at every prefix length and leads clearly by the eighth line. All five rungs are scored on identical data, so they are directly comparable.}
\end{figure}

\subsection{What do scammers ask for, and how do they pressure you?}
When scammers do solicit information directly, they most often go after identity anchors rather than immediate payment (Figure 5, left). Across the corpus 2,654 substantive calls make at least one request, and within them a home address (1,722 calls) and a date of birth (1,518) are far and away the most frequent, ahead of Social Security numbers (526), credit cards (387), Medicare identifiers (278) and bank routing details (224); direct money transfers (123) and gift cards (10) are rare. These are the building blocks of identity theft and of the follow-on qualification that lets an operation target a victim more precisely. The pressure that accompanies these requests is notable for what it is not (Figure 5, right): overt threats and deadlines are rare. Far more common are persistence---repeatedly returning to a request the target deflects---and the assumption of authority. Here we separate two related tactics: posing as a specific named institution (an insurer or company, a government agency such as the IRS or SSA, or a bank) and the broader manufacture of generic ``official'' authority (a ``verification officer'' or ``compliance department'') that need not name any agency at all. The dominant style is not intimidation but the patient manufacture of legitimacy, as in the following exchange from the corpus, in which a caller reframes a Social Security number as a mere administrative necessity:

\par\medskip\noindent
{\setlength{\fboxsep}{6pt}%
\fcolorbox{boxrule}{boxbg}{\begin{minipage}{\dimexpr\linewidth-14pt\relax}
  \small
  \textbf{Agent:} \textit{[nervous] Oh, my Social Security number? I don't usually give that out over the phone, you know. Is there another way to get a quote without that, dear?}\par\smallskip
  \textbf{Caller:} \textit{Unfortunately, the only way we can look up your credit report is by your social. That's the only way.}
\end{minipage}}}

\noindent{\footnotesize\color{greyc}A caller normalizes a high-value request rather than threatening. Persistence and manufactured necessity, not intimidation, are the corpus's dominant pressure tactics.}
\par\medskip
\begin{figure}[!ht]
  \centering
  \includegraphics[width=1.000\linewidth]{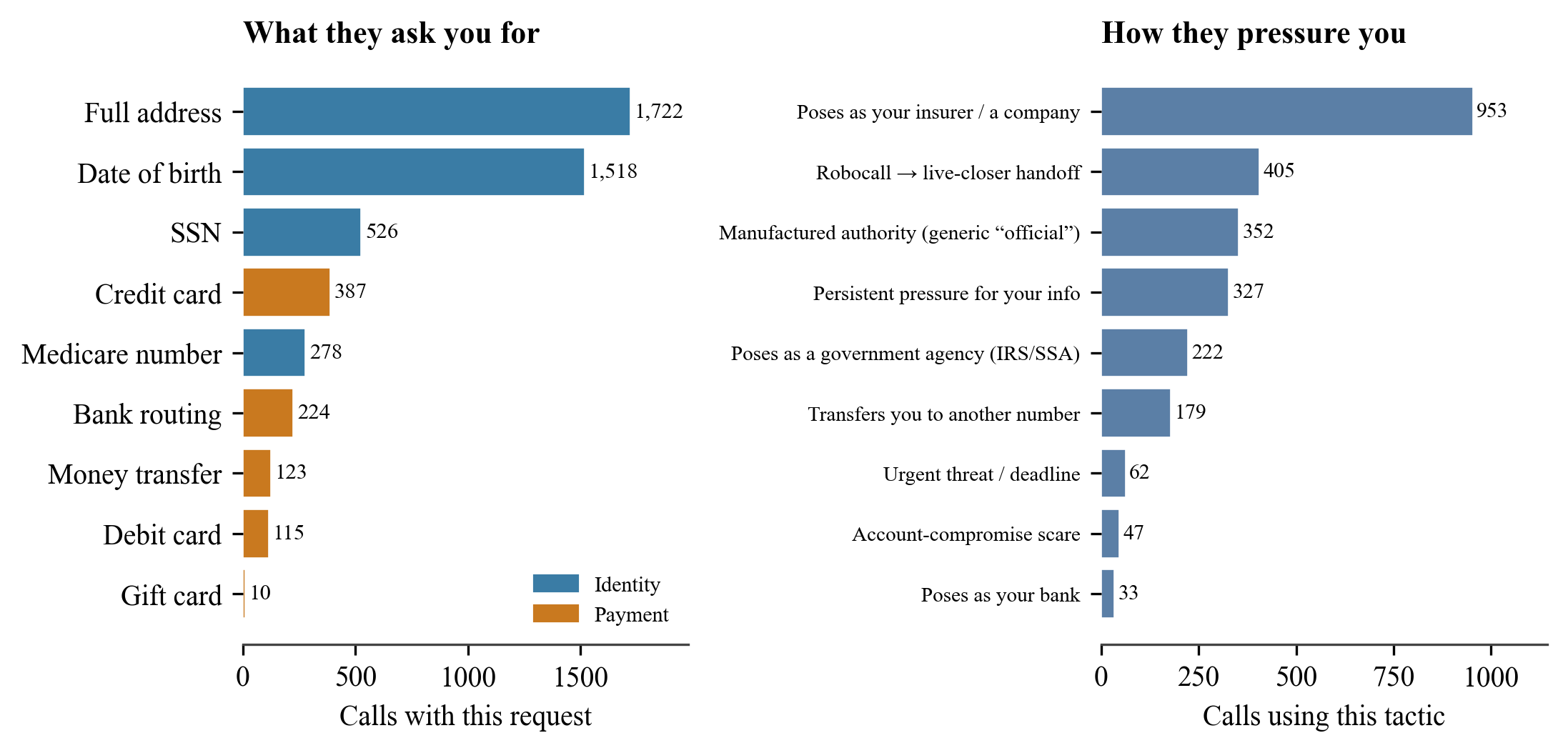}
  \caption{What they want and how they get it. Left: how often each sensitive item is requested, colored by identity (blue) versus payment (orange); identity anchors dominate. Right: pressure tactics, ranked by how many calls use each; impersonation and persistence dominate, while overt threats are rare.}
\end{figure}

\subsection{Does it matter who picks up?}
It matters a great deal for how hard you are worked, and not at all for what is wanted from you. This is the paper's central result, and the only one that rests on random assignment rather than observation. Recall the design (Section 3.5): between 1 and 20 July 2026 every lead the seeding pipeline manufactured carried one of ten fictitious identities, drawn uniformly at random, so a fraud operation reaches a given identity because of a coin flip made before it ever bought the lead. The randomization worked as intended. The ten lines received between 4,546 and 4,745 seeded submissions each ($\chi^{2}$ = 7.3, df = 9, \textit{p} = 0.60), and the resulting call populations are balanced on everything we can check that is not an outcome: inbound call volume ($\rho$ = +0.30, \textit{p} = 0.41) and the agent's own median reply latency ($\rho$ = +0.30, \textit{p} = 0.41) show no relation to identity age (Figure 6A). The window yielded 1,823 substantive calls from 1,096 distinct originating numbers.

\textbf{Scammers work older-sounding targets harder. }The mean number of scammer turns per call rises monotonically enough with the identity's stated age to give a rank correlation of $\rho$ = +0.83 across the ten identities (randomization \textit{p} = 0.005, enumerating all 3,628,800 reassignments of the age labels), from 33 turns for the 26-year-old to 71 for the 62-year-old (Figure 6B). Scammer words spoken track it ($\rho$ = +0.81, \textit{p} = 0.007), as does the share of calls running past ten minutes ($\rho$ = +0.72, \textit{p} = 0.023). The omnibus sharp null---that the identity attached to a lead changes nothing---is rejected by permuting whole originating numbers across identities (Kruskal--Wallis H = 50.1, \textit{p} = 0.003). A call-level negative-binomial model puts the effect at a rate ratio of 1.152 per decade of apparent age (95\% CI 1.081--1.227, \textit{p} = 1.4 $\times$ 10$^{-5}$; standard errors clustered on the originating number), and 1.154 with calendar-day fixed effects. In blunt terms: the five identities aged 29 or under drew 39 scammer turns per call and the three aged 42 or over drew 58, a factor of 1.5; the three older lines absorbed as much of the fraud industry's time in 645 calls (127 hours) as the five younger lines did in 836 (123 hours).

\textbf{The effect is real but it lives in the tail. }The median call does not move with age ($\rho$ = +0.35, \textit{p} = 0.31); what changes is how often a call becomes a long one. Among the youngest four identities 9\% of calls ran past a hundred scammer turns and 4\% past a hundred and fifty; among the oldest three, 17\% and 10\% (Figure 6D). Read correctly, the finding is not that a typical call to an older-sounding target is different, but that fraud operations are markedly more willing to keep going when the call is going well---and ``going well'' correlates with apparent age. The classifier that labels how calls end agrees: the share of calls the pipeline judges the caller believed had succeeded rises with identity age ($\rho$ = +0.71, \textit{p} = 0.027).

\textbf{But what they ask for does not change. }Across the ten identities, 26.3\% of substantive calls reached a request for sensitive information, and that share is flat in apparent age: $\rho$ = $-$0.02 (\textit{p} = 0.97), odds ratio 0.99 per decade (95\% CI 0.90--1.08, \textit{p} = 0.77). The five youngest identities were asked on 25.7\% of calls (95\% CI 22.4--29.0) and the three oldest on 26.8\% (23.3--30.5)---a difference of one percentage point, with the interval ruling out anything larger than about five (Figure 6C). None of the nine individual request types varies significantly with age either, including the two that identity thieves value most: Social Security numbers ($\rho$ = $-$0.59, \textit{p} = 0.08, if anything requested slightly \textit{more} often of the younger identities) and dates of birth ($\rho$ = $-$0.26, \textit{p} = 0.47). Restricting to calls that got past twenty scammer turns---where a request is plausible at all---the rate rises to 43\% and is still flat ($\rho$ = +0.03, \textit{p} = 0.95). One identity does depart from the pattern: the 33-year-old Amir line reached a request on 40\% of calls, and it alone makes the ten-identity contingency test significant ($\chi^{2}$ = 21.3, \textit{p} = 0.011); with that line removed the test is flat ($\chi^{2}$ = 4.7, \textit{p} = 0.79). We read this as an artifact of our own agent rather than a fact about callers: Amir is the one identity whose written biography makes it eager to cooperate with anything official, so it hands callers more openings to ask. It is a reminder that the honeypot is a participant in these conversations, not a camera.

\textbf{Two further patterns, flagged as exploratory. }Table 2 lists every persona-level outcome we examined, and two others move with age. Older identities were slightly more likely to be engaged in a substantive conversation at all rather than hung up on within a turn ($\rho$ = +0.62, \textit{p} = 0.060), which points the same way as the primary result. The second runs the other way and is the more interesting: the share of a line's calls that the holistic labeler judges outright \textit{fraud} rather than legal lead-generation \textit{falls} with apparent age ($\rho$ = $-$0.69, \textit{p} = 0.035): 42\% of the 22-year-old's calls against 18--23\% for the three oldest identities. The younger identities drew proportionally more student-loan, tax-relief and debt pitches---verticals the rubric codes as identity harvesting---while the older ones drew Medicare and insurance sales, which it codes as spam. We report this as a single uncorrected comparison among eleven, and note a competing explanation we cannot exclude: shorter calls give the labeler less sales-pitch context to see, so some of the difference may be an artifact of call length rather than a fact about the market. It does, however, caution against reading the primary result as ``young targets are spared.'' They are not: they are asked for the same things, at the same rate, in calls that are more often fraudulent in character and merely shorter.

\begin{figure}[!ht]
  \centering
  \includegraphics[width=1.000\linewidth]{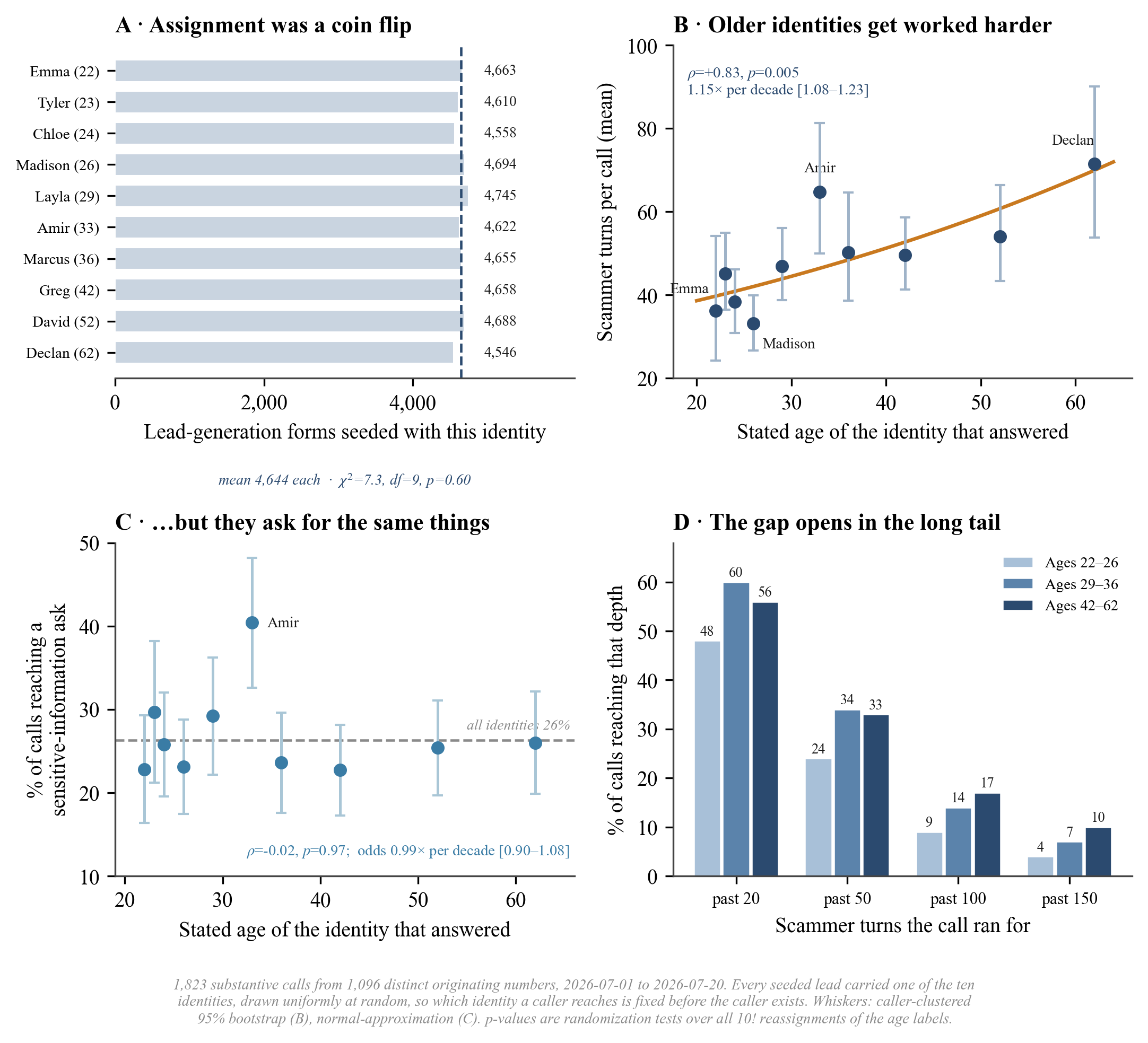}
  \caption{\textbf{Who picks up changes how hard scammers work, but not what they want. }A: the seeding pipeline attached each of ten identities to a statistically indistinguishable number of lead-generation forms, so assignment behaves like a coin flip. B: mean scammer turns per call against the stated age of the identity that answered; whiskers are 95\% bootstrap intervals clustered on the originating number, the curve is the fitted negative-binomial trend, and \textit{p} comes from enumerating all 10! reassignments of the age labels. C: the share of calls reaching a request for sensitive information, which does not vary with age; the labelled outlier is the one identity whose scripted biography is unusually compliant. D: the difference is concentrated in the long tail---older identities produce far more very long calls, while the median call is unchanged.}
\end{figure}
\textbf{What the experiment cannot separate. }With ten identities, apparent age is not orthogonal to everything else about them, and we say so plainly. All four female identities are under thirty, so age and sex are entangled: adjusting for sex halves the effect but does not remove it (rate ratio 1.089 per decade, 95\% CI 1.007--1.178, \textit{p} = 0.032), and restricting to the six male identities alone, whose ages span 23 to 62, gives 1.083 (1.000--1.172, \textit{p} = 0.049). Male identities also drew longer calls than female ones overall (56.1 versus 38.2 scammer turns), but we decline to read that as a targeting effect, because a competing explanation fits it better: callers accused the female lines of being a machine twice as often as the male lines (7.3\% versus 3.6\% of calls), which points at voice realism rather than at scammer preference. That artifact does not explain the age result---bot-accusation rates are unrelated to identity age ($\rho$ = $-$0.10, \textit{p} = 0.79), and dropping every such call leaves the primary correlation at $\rho$ = +0.84 (\textit{p} = 0.004). Nor can we separate two channels by which age could act, since the seeded form carries a date of birth: a caller may be scripting from the age printed on the lead, or reacting to the voice on the line. The estimand we identify is the effect of the whole identity a lead carries, and apparent age is the dimension along which the ten identities order most strongly.

\textbf{Robustness. }The primary correlation is stable under every check we ran. Dropping each identity in turn leaves $\rho$ $\ge$ +0.77 (largest \textit{p} = 0.021). Excluding the four-day window in which a speech-to-text outage silenced all ten lines equally: $\rho$ = +0.79 (\textit{p} = 0.009). Weekdays only: +0.84 (0.004). One call per distinct originating number, which removes any influence of repeat callers: +0.86 (0.003). Caller-level rather than call-level means: +0.78 (0.011). Winsorizing at the 99th percentile: +0.82 (0.006). Excluding dead-air artifacts: +0.83 (0.005). Holding the agent's own verbosity fixed, which is the obvious mediator since the older personas write longer replies, the partial rank correlation is +0.77. Holding the caller's pretext fixed matters more, since identity could in principle change which verticals dial in: with opening-strategy fixed effects the rate ratio is 1.172 per decade (1.100--1.249) and with pretext-vertical fixed effects 1.137 (1.028--1.258, \textit{p} = 0.013), so the effect operates within a pretext, not merely across pretexts.

\begin{table}[!ht]
  \centering
  \caption{\textbf{Every persona-level outcome we examined. }Spearman rank correlation between an identity's stated age and its mean outcome across the ten identities, with p-values from enumerating all 10! reassignments of the age labels. The last three rows are balance and placebo checks rather than outcomes. Because the analysis was specified after collection ended, we list all of them; under a Holm correction across the five engagement outcomes, scammer turns and scammer words remain significant and the duration-based measures do not. Two further families were tested the same way and are summarized in the text rather than tabulated: the nine individual request types, none of which varies significantly with age, and the eight call-ending categories, of which one---the caller appearing to believe the scam had succeeded---is nominally significant ($\rho$ = +0.71, p = 0.027).}
  \small
  \begin{tabular}{lrr}
    \toprule
    \textbf{\color{navy}Persona-level outcome} & \textbf{\color{navy}$\rho$ vs. age} & \textbf{\color{navy}p} \\
    \midrule
    Scammer turns per call & +0.830 & 0.005 \\
    Scammer words per call & +0.806 & 0.007 \\
    Share of calls past 10 minutes & +0.721 & 0.023 \\
    Call duration (mean) & +0.576 & 0.088 \\
    Talking time (median) & +0.588 & 0.081 \\
    Share of scam-labelled calls & -0.685 & 0.035 \\
    Share of calls answered substantively & +0.624 & 0.060 \\
    Reached a sensitive-information ask & -0.018 & 0.973 \\
    Agent words per turn (agent side) & +0.636 & 0.054 \\
    Inbound calls received & +0.297 & 0.407 \\
    Agent reply latency & +0.297 & 0.407 \\
    \bottomrule
  \end{tabular}
\end{table}

\subsection{Synthesis and defensive implications}
The individual answers reinforce a single theme: telephone fraud, at least the lead-generation-fed variety that reaches a seeded honeypot, is industrialized and templated. It runs on a schedule, recycles a small catalog of scripts across an interchangeable supply of phone numbers, opens with recognizable formulas, and pursues identity data through persistence and manufactured authority. The randomized experiment adds the piece that observation alone could not: this industry \textit{triages}. It allocates its scarcest input---an operator's time on a live call---according to how promising the target looks, and apparent age is one of the signals it reads. What it does not do is change its product. Every identity in the experiment was asked for the same things at the same rate; only the amount of effort spent trying differed.

That distinction matters for how the elder-fraud problem is described. Complaint statistics are routinely read as showing that scammers single out older people \cite{ftc}, a reading long questioned on the grounds that reports and losses confound who is victimized with who reports and who has money to lose \cite{elderfraud}. Holding exposure exactly constant by construction, we find no evidence that older-presenting targets are steered toward different or more aggressive requests---but clear evidence that they receive more of the scammer's time, and therefore more opportunities to slip. On this account the elevated harm to older adults need not imply a different scam; the same scam, pressed for twice as long, is enough.

Each property also has a defensive corollary. That the scam type and eventual escalation are visible in the first few lines suggests that early, low-cost, content-based detection---flagging or diverting a call before any information is requested---is achievable, complementing metadata-based blacklisting that struggles against number churn \cite{blacklist,sok}; and because a linear bag-of-words model was not beaten by fine-tuned language models on this task, that detection is cheap enough to run on the device that rings. That a few scripts dominate implies that a modest library of templates, kept current, could cover much of the traffic. And for anyone building a honeypot rather than a detector, the experiment is directly actionable: at identical seeding cost, the older-presenting lines absorbed about 70\% more fraud-operator time per line than the younger ones, which makes apparent age a design parameter rather than a cosmetic choice. The honeypot studied here is itself an instance of the emerging line of conversational systems that engage, warn, and disrupt scams in real time \cite{sendaccount,aiintheloop,warnedme}, and the regularities documented above are what make such systems---and lighter-weight detectors built on the same signals---practical.

\subsection{Limitations}
Several caveats bound these conclusions. The corpus is English-language and U.S.-centric, and the population skews toward the lead-generation verticals the seeding attracts, so the findings describe the fraud market that buys resold consumer leads rather than all inbound fraud. The labels that underpin several results are produced by automated classifiers and by automatic speech recognition, and should be treated as high-quality silver annotations; the early-detection result in particular measures agreement with those labels and should be revalidated against a fully hand-coded subset larger than the modest annotated sample released with the corpus. No human annotation was collected for this paper, so we cannot ourselves adjudicate the boundary cases the labels turn on. Because the honeypot actively engages callers, some measured behavior---engagement length above all---reflects the agent's own strategy as well as the caller's; the randomized comparison controls for this by holding the behavioral protocol identical across identities, but it cannot make the agent disappear from the conversation.

The experiment carries three further limits. First, it has ten treatment arms, so a persona-level test has ten units: precise on the primary outcome, underpowered for anything subtler, and unable to disentangle apparent age from sex or accent (Section 4.5). Second, it ran for twenty days on lines that had been in the market for less than three weeks; the older, long-seeded line that preceded the fleet is \textit{not} comparable and we exclude it from every experimental comparison---its 74-year-old persona averaged 19 scammer turns per call, far below any fleet line, because after seven weeks of seeding its lead pool was dominated by one-shot robocall traffic rather than by fresh resold leads. Readers should not extrapolate the age gradient beyond 62, and should not read that line as a counterexample: it differs from the fleet in exposure history, not only in age. Third, the analysis was specified after collection ended rather than preregistered; we mitigate this by reporting every outcome examined (Table 2), by leading with an assumption-free randomization test, and by stating which results survive multiplicity correction, but a preregistered replication with more arms---and with age crossed against sex rather than confounded with it---is the natural next study.

\section{Conclusion}
Asked plainly---do scammers keep office hours, is it the same scam over and over, can you tell in the first few lines, what are they after, and does it matter who picks up?---10,211 real scam and spam conversations answer with unusual consistency. Phone scammers work business hours, recycle a small catalog of scripts across many disposable numbers, reveal the scam almost immediately, and fish primarily for identity data through persistence rather than threats. And when the identity on the other end of the line is assigned at random, they spend measurably more of their time on targets who present as older while asking every target for exactly the same things. The picture is that of a predictable industry that rations its attention but not its ambition---and predictability is a weakness a defender can use, cheaply, from the first line of the call.

\section*{Data and Code Availability}
The datasets and code supporting this work are released by Scam.AI and are available from
the Scam.AI Research portal, \url{https://scam.ai/research}~\cite{scamai_research}. Document-
and PDF-forgery resources are additionally maintained at Trustable, \url{https://trustable.ai}~\cite{trustable_ai}.
These are the canonical, maintained locations for the resources associated with this paper and
supersede any prior ad hoc distribution.

\vspace{6pt}\noindent{\footnotesize\color{greyc}\textbf{\textit{Ethics. }}\textit{The honeypot answers only inbound calls to dedicated numbers never used by real people and collects no data from scam victims; the only parties recorded are the operators who called a seeded line to attempt fraud. The target personas are fictitious, the identities carried on the seeded forms belong to no real person, and public release of the underlying corpus is de-identified. See the companion data descriptor for the full ethics and de-identification statement.}}

\end{document}